\documentstyle[prl,aps,epsfig]{revtex}

\begin{document}

\draft

\title{Evidence for a Nodeless Gap from the Superfluid Density
of Optimally Doped Pr$_{1.855}$Ce$_{0.145}$CuO$_{4-y}$ Films}

\author{John A. Skinta and Thomas R. Lemberger}
\address{Department of Physics, Ohio State University, Columbus,
OH 43210-1106}
\author{T. Greibe and M. Naito}
\address{NTT Basic Research Laboratories, 3-1 Morinosato Wakamiya,
Atsugi-shi, Kanagawa 243, Japan}

\date{\today}
\maketitle

\begin{abstract}

We present measurements of the \textit{ab}-plane magnetic
penetration depth, $\lambda(T)$, in five optimally doped
Pr$_{1.855}$Ce$_{0.145}$CuO$_{4-y}$ films for $1.6$ K $\leq T \leq
T_c \sim 24$ K. Low resistivities, high superfluid densities
$n_s(T) \propto \lambda^{-2}(T)$, high $T_c$'s, and small
transition widths are reproducible and indicative of excellent
film quality. For all five films,
$\lambda^{-2}(T)/\lambda^{-2}(0)$ at low $T$ is well fitted by an
exponential temperature dependence with a gap, $\Delta_{min}$, of
0.85 $k_B T_c$. This behavior is consistent with a nodeless gap
and is incompatible with \textit{d}-wave superconductivity.

\end{abstract}


It is widely accepted that pairing symmetry in the hole-doped
cuprates is predominantly $d_{x^2-y^2}$, at least near optimal
doping where most phase-sensitive measurements have been made
\cite{harlingen,tsuei01}. Essentially all other experimental
results agree with this view. The developing situation in the
electron-doped cuprates is not so clear. While recent
phase-sensitive measurements on optimally doped
Nd$_{1.85}$Ce$_{0.15}$CuO$_{4-y}$ (NCCO) and
Pr$_{1.85}$Ce$_{0.15}$CuO$_{4-y}$ (PCCO) films are consistent with
a $d_{x^{2} - y^{2}}$ energy gap \cite{tsuei02}, there is
substantial evidence for a nodeless gap. The penetration depth,
$\lambda(T)$, measured by Alff \textit{et al}. \cite{alff01} via
field modulation of Josephson junctions fabricated in the
\textit{ab}-plane of optimally doped PCCO films is exponentially
flat at low $T$, suggesting a gapped superconducting state. The
zero-bias peak seen in tunnelling data from hole-doped cuprates --
associated with Andreev bound states of a \textit{d}-wave order
parameter -- is absent from tunnel junctions in NCCO
\cite{alff02,kashiwaya}, suggesting \textit{s}-wave
superconductivity. In light of persuasive evidence for $d_{x^{2} -
y^{2}}$ superconductivity from phase-sensitive \cite{tsuei02} and
angle resolved photoemission spectroscopy \cite{armitage}
measurements on electron-doped samples, one might suspect some
problem with measurements on in-plane tunnel junctions, or with
the films themselves. This suspicion may be strengthened by
reports of quadratic behavior of $\lambda(T) - \lambda(0)$ at low
$T$ in optimally doped PCCO crystals \cite{kokales01,prozorov}.
Thus the importance of penetration depth measurements that confirm
the findings of Alff \textit{et al}. and demonstrate high sample
quality and sample-to-sample reproducibility.

We present measurements of the \textit{ab}-plane superfluid
density, $n_s(T) \propto \lambda^{-2}(T)$, in five optimally doped
Pr$_{1.855}$Ce$_{0.145}$CuO$_{4-y}$ films. $\lambda^{-2}(T)$ at
low $T$ in hole-doped cuprates is consistently linear
\cite{kamal,lee} or quadratic \cite{bonn} in temperature. Theory
has not found a scenario in which behavior flatter than $T^2$ is
predicted for $d_{x^{2} - y^{2}}$ superconductors
\cite{hirschfeld,annett,kosztin,won}. On the other hand,
$\lambda^{-2}(T)$ in gapped -- e.g., \textit{s}-wave --
superconductors is exponentially flat at low temperatures. As
shown below, $\lambda^{-2}(T)$ in our optimally doped PCCO films
is exponential in $\Delta_{min}/T$, with the \emph{same} value of
the minimum gap, $\Delta_{min} = 0.85 \pm 0.05$ $k_B T_c$.

Our films were prepared by molecular-beam epitaxy on 12.7 mm
$\times$ 12.7 mm $\times$ 0.35 mm SrTiO$_{3}$ substrates as
detailed elsewhere \cite{naito}. Ce content, $x$, is measured
using inductively coupled plasma spectroscopy and is known to
better than $\pm 0.005$. Films are highly oriented with their
\textit{c}-axes perpendicular to the substrate. Table I summarizes
film properties. Thicknesses vary from 750 \AA \space to 1250 \AA,
and one film (P6) was grown on a 250 \AA \space buffer layer of
insulating Pr$_2$CuO$_4$. \textit{ab}-plane resistivities,
$\rho(T)$ (Fig. 1), are reproducible to $\pm 15\%$. $T_c = 23.7
\pm 0.5$ K and transition width, $\Delta T_c = 0.9 \pm 0.2$ K meet
or exceed values reported for nominally identical crystals
\cite{kokales01,prozorov}. As noted previously \cite{naito},
residual resistivities of PCCO films are smaller than those of
PCCO crystals \cite{kokales02}. Curvature in $\lambda(T) -
\lambda(0)$ varies by more than a factor of two from crystal to
crystal \cite{kokales01,prozorov}. These comparisons suggest that
film quality is superior to crystal quality at this time.
$\rho(T)$ in our films at 40 K is a factor of two lower than that
of high-quality, comparably doped crystals of
La$_{2-x}$Sr$_x$CuO$_4$ (LSCO) \cite{ando}, the hole-doped cousin
of PCCO. $\lambda(0) = 1800 \pm 300$ \AA \space in our films
matches the lowest value, 1930 \AA, reported in LSCO near optimal
doping \cite{panagopoulos}, and it is only slightly larger than
the \textit{a}-axis penetration depth, 1600 \AA, in YBCO
\cite{basov}. We therefore conclude that our films have very
little disorder.

We measure $\lambda^{-2}(T)$ with a low frequency two-coil mutual
inductance technique described in detail elsewhere
\cite{turneaure01}. A film is centered between two small coils,
and a current at about 50 kHz in one coil induces eddy currents in
the film. Currents are approximately uniform through the film
thickness. Data have been measured to be independent of frequency
for $10$ kHz $\leq f \leq 100$ kHz. Magnetic fields from the
primary coil and the film are measured as a voltage across the
secondary coil. We have checked that the typical excitation field
(100 $\mu$Tesla $\perp$ to film) is too small to create vortices
in the film. Because the coils are much smaller than the film, the
applied field is concentrated near the film's center and
demagnetizing effects at the film perimeter are irrelevant. All
data presented here are in the linear response regime.

Because film thicknesses, $d$, are less than $\lambda$, it is the
sheet conductivity $\sigma(\omega,T)d = \sigma _{1}(\omega,T)d -
i\sigma_{2}(\omega,T)d$ that is measured, with an estimated
accuracy of 5\%. $\lambda^{-2}(T)$ is obtained from $\sigma_2$:
$\lambda ^{-2}(T) \equiv \mu _{0} \omega \sigma_{2}(T)$, where
$\mu _{0}$ is the magnetic permeability of vacuum, and accuracy is
limited by 5\% uncertainty in $d$. The temperature dependence of
$\lambda ^{-2}(T)/\lambda ^{-2}(0)$ is unaffected by uncertainty
in $d$. The origin of 0.2\% ``wiggles'' in $\lambda^{-2}(T)$ at
low temperatures is uncertain, but we know that at least some of
the effect arises from slow drift in amplifier gain.

$T_{c}$ and  $\Delta T_c$ (Table I) are defined to be the position
and full-width of the fluctuation peak in $\sigma_{1}(T)$ measured
at 50 kHz (Fig. 2). $T_c$ from resistivity (Fig. 1) and
penetration depth measurements (Figs. 2 and 3) are identical.
Structure in $\sigma_1(T)$ is due to slight variation in $T_c$
through the film thickness. Our measurement technique reveals
transitions in all layers of the sample, so $\Delta T_c \approx 1$
K indicates excellent film homogeneity.

Figure 3 displays $\lambda^{-2}(T)$ in all five films (thick solid
lines). The spread in $\lambda^{-2}(0)$ is larger than expected
from uncertainty in $d$ \cite{note2}. Slight upward curvature in
$\lambda^{-2}$ near $T_c$ is reproducible. Most importantly, the
flatness of the low-temperature data is highly reproducible.

To determine the behavior of $\lambda^{-2}(T)$ at low
temperatures, we fit the first $\sim 5$\% drop in
$\lambda^{-2}(T)/\lambda^{-2}(0)$ to
\begin{equation}
\label{eq:02} \lambda^{-2}(T)/\lambda^{-2}(0) \sim  1 -
C_{\infty}e^{-D/t},
\end{equation}
with $t \equiv T/T_c$. $D$ is roughly the minimum gap value,
$\Delta_{min}$, normalized to $T_c$ and is fixed at 0.85 for all
films to emphasize sample-to-sample reproducibility. $C_{\infty}$
and $\lambda^{-2}(0)$ (Table I) are adjusted for each film to
yield the fits (dotted curves) in Fig. 4. The exponential curves
in Fig. 4 \emph{are not best fits}, but their $\chi^2$ values
(Table I) lie within the estimated experimental noise of $\pm
0.2$\% (error bars in Fig. 4). The worst fits are improved by a
factor of two by adjusting $D$ by 10\%. We note that a more
thorough analysis finds that the peak in the density of states is
near 2.2 $k_B T_c$ in optimally doped PCCO \cite{skinta}.

For comparison, quadratic and cubic fits of the form
$\lambda^{-2}(T) \sim \lambda^{-2}(0) \left[ 1 - \left( T/T_0
\right)^x \right]$ with $x=2$ or 3, were also performed. They are
shown for film P2 in Fig. 4. For all films, the best quadratic
fits are visibly poorer than cubic or exponential fits. $\chi^2$
values for quadratic fits are typically twice as large as for
exponential fits. We note that cubic fits possess the smallest
$\chi^2$ values, leaving open the possibility that the
superconducting density of states increases as energy cubed at low
energies rather than being fully gapped \cite{skinta}. No
\textit{d}-wave theory predicts such behavior.

Figure 3 displays $\lambda^{-2}(T)$ from 0 K to $T_c$ along with
low-$T$ exponential fits from Fig. 4 and low-$T$ quadratic and
cubic fits extrapolated to higher temperatures. Exponential fits
are cut off at 8 K, since the temperature dependence of the gap is
neglected in the low-$T$ expression, Eq. (1). We have already
demonstrated that low-$T$ quadratic fits are unacceptably poor.
Even if one believes that the low-temperature data are quadratic
in $T$, there is no sign of a crossover to linear behavior
\cite{hirschfeld}, and the extrapolated low-$T$ quadratics go to
zero well above $T_c$, so the curvature is too weak to be
compatible with a dirty \textit{d}-wave interpretation
\cite{hirschfeld}. We note that previous observations of quadratic
behavior in $\lambda(T)$ \cite{kokales01,prozorov} do not find a
crossover from quadratic to linear, weakening their support for
\textit{d}-wave superconductivity.

In conclusion, we have presented high-precision measurements of
$\lambda^{-2}(T)$ in five optimally doped PCCO films.
Resistivities, $T_c$'s, transition widths, and superfluid
densities are reproducible and indicative of superior film
quality. In every film, $\lambda^{-2}(T)$ at low temperatures is
flatter than $T^2$ and is well fitted by an exponential
temperature dependence with the same minimum gap value of 0.85
$k_B T_c$, indicating a nodeless gap in optimally doped PCCO.

%\pagebreak

%Figure 1
\begin{figure}[htb]
\centerline{\epsfxsize=3.0in \epsfbox{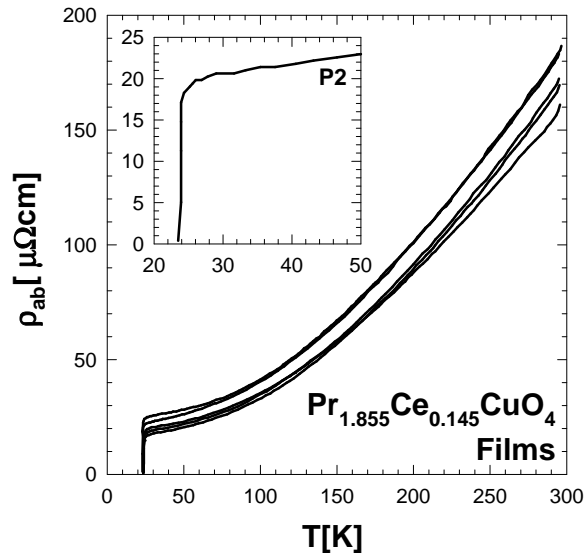}} \vspace{0.1in}
\caption{\textit{ab}-plane resistivities, $\rho(T)$, of five
optimally doped PCCO films. Inset displays transition region for a
typical film, P2.} \vspace{-0.1in}
\end{figure}

%Figure 2
\begin{figure}[htb]
\centerline{\epsfxsize=3.0in \epsfbox{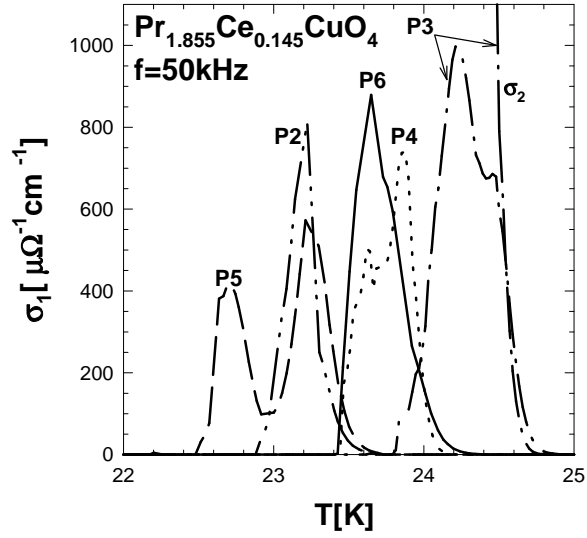}} \vspace{0.1in}
\caption{$\sigma_1(T)$ at 50 kHz for five optimally doped PCCO
films. $T_c$ and $\Delta T_c$ are the temperature and full-width
of the fluctuation peak in $\sigma_1$. For comparison,
$\sigma_2(T)$ is shown for film P3.} \vspace{-0.1in}
\end{figure}

%Figure 3
\begin{figure}[htb]
\centerline{\epsfxsize=3.0in \epsfbox{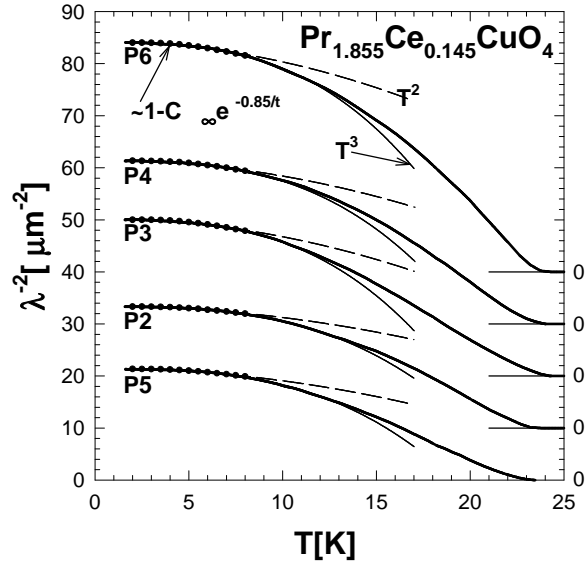}} \vspace{0.1in}
\caption{$\lambda ^{-2}(T)$ (thick lines) for the
\textit{ab}-plane of five optimally doped PCCO films. Successive
curves are offset by 10 $\mu$m$^{-2}$ for clarity. Thin solid
(dashed) lines are extrapolations of the best cubic (quadratic)
fits to low-$T$ data. Dotted curves are exponential fits to
low-$T$ data, and $t \equiv T/T_c$. See Fig. 4 and text for
explanation of low-$T$ fits.} \vspace{-0.1in}
\end{figure}

%Figure 4
\begin{figure}[htb]
\centerline{\epsfxsize=3.0in \epsfbox{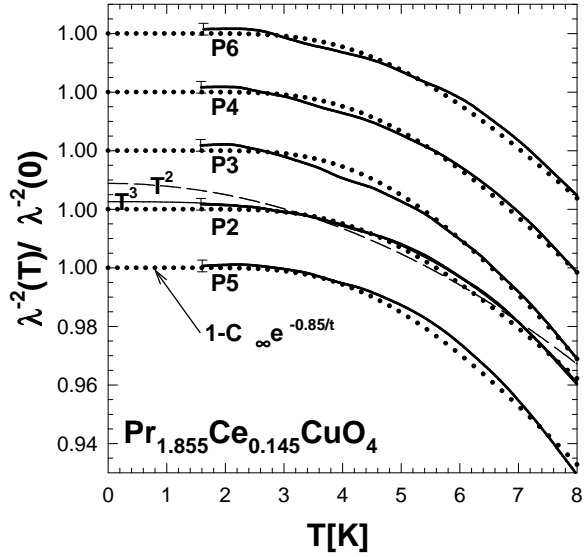}} \vspace{0.1in}
\caption{First $\sim 5$\% drop in $\lambda ^{-2}(T)/\lambda
^{-2}(0)$ for five optimally doped PCCO films. Successive curves
are offset by 0.02 for clarity. Dotted curves are fits to this
data of $ 1 - C_{\infty}e^{-D/t}$, with $t \equiv T/T_c$ and
$D=0.85$. The fits lie within the experimental noise of 0.2\%,
represented by error bars. The thin solid (dashed) line is a cubic
(quadratic) fit, described in the text. Best-fit quadratics lie
outside the experimental noise and are therefore unacceptable.}
\vspace{-0.1in}
\end{figure}

%Table I
\begin{table}
\caption{PCCO film properties. $d$ is film thickness. $T_{c}$ and
$\Delta T_{c}$ are the temperature and full-width of the
fluctuation peak in $\sigma_{1}(T)$. Uncertainty in $d$ and
$\lambda^{-2}(0)$ are estimated to be $\pm 5$\%. $\rho$(30 K) is
the \textit{ab}-plane resistivity just above $T_c$. $C_{\infty}$
is a fit parameter from Eq. (1). $\chi^2 \equiv (1/N)\sum{[\lambda
^{-2}(T) - \lambda_{fit}^{-2}(T)]^2/[\lambda_{fit}^{-2}(T)]^2}$,
where $N$ = number of data points, is a measure of exponential fit
quality; $\chi^2$ from $T^2$ fits are in every case poorer.}
\label{table01}
\begin{tabular}{c c c c c c c c}
Film & $d$ [\AA] & $T_c$ [K] & $\Delta T_c$ [K] &
$\lambda (0)$ [\AA] & $\rho$(30 K) [$\mu \Omega$ cm] & $C_{\infty}$ & $\chi^2$ [$\times 10^{-6}$] \\
\hline \hline
P2 & 750  & 23.2 & 0.7 & 2100 & 21 & 0.68 & 2.56 \\
P3 & 1000 & 24.2 & 1.0 & 1800 & 19 & 0.93 & 3.69 \\
P4 & 1250 & 23.9 & 0.7 & 1800 & 18 & 0.78 & 2.30 \\
P5 & 750  & 23.2 & 1.1 & 2200 & 23 & 0.79 & 4.86 \\
P6 & 750  & 23.6 & 0.8 & 1500 & 26 & 0.69 & 3.22 \\
%\hline
\end{tabular}
\end{table}


\begin{references}

\bibitem{harlingen} D.J. Van Harlingen, Rev. Mod. Phys. {\bf 67},
515 (1995).
\bibitem{tsuei01} C.C. Tsuei and J.R. Kirtley, Rev. Mod. Phys.
{\bf 72}, 969 (2000).
\bibitem{tsuei02} C.C. Tsuei and J.R. Kirtley, Phys. Rev. Lett.
{\bf 85}, 182 (2000).
\bibitem{alff01} L. Alff \textit{et al}., Phys. Rev. Lett. {\bf 83},
2644 (1999).
\bibitem{alff02} L. Alff \textit{et al}., Phys. Rev. B {\bf
58}, 11197 (1998).
\bibitem{kashiwaya} S. Kashiwaya \textit{et al}., Phys. Rev. B
{\bf 57}, 8680 (1998).
\bibitem{armitage} N.P. Armitage \textit{et al}., Phys. Rev. Lett.
{\bf 86}, 1126 (2001).
\bibitem{kokales01} J.D. Kokales \textit{et al}., Phys. Rev. Lett.
{\bf 85}, 3696 (2000).
\bibitem{prozorov} R. Prozorov \textit{et al}., Phys. Rev. Lett.
{\bf 85}, 3700 (2000).
\bibitem{kamal} S. Kamal \textit{et al}., Phys. Rev. B {\bf 58},
8933 (1998).
\bibitem{lee} S.-F. Lee \textit{et al}., Phys. Rev. Lett. {\bf 77},
735 (1996).
\bibitem{bonn} D.A. Bonn \textit{et al}., Phys. Rev. B {\bf 50},
4051 (1994).
\bibitem{hirschfeld} P.J. Hirschfeld and N. Goldenfeld, Phys.
Rev. B {\bf 48}, 4219 (1993).
\bibitem{annett} J.F. Annett, N. Goldenfeld, and S.R. Renn, in
{\it Physical Properties of High Temperature Superconductors II},
edited by D.M. Ginsberg (World Scientific, Singapore, 1990).
\bibitem{kosztin} I. Kosztin and A.J. Leggett, Phys. Rev. Lett.
{\bf 79}, 135 (1997).
\bibitem{won} H. Won and K. Maki, Phys. Rev. B {\bf 49}, 1397
(1994).
\bibitem{naito} H. Yamamoto, M. Naito, and H. Sato, Phys. Rev. B
{\bf 56}, 2852 (1997); M. Naito, H. Sato, and H. Yamamoto, Physica
(Amsterdam) {\bf 293C}, 36 (1997).
\bibitem{kokales02} J.D. Kokales \textit{et al}., Physica (Amsterdam) {\bf
341-348C}, 1655 (2000).
\bibitem{ando} Y. Ando, A.N. Lavrov, S. Komiya, K. Segawa, and
X.F. Sun, Phys. Rev. Lett. {\bf 87}, 017001 (2001).
\bibitem{panagopoulos} C. Panagopoulos \textit{et al}., Phys. Rev. B {\bf 60},
14617 (1999).
\bibitem{basov} D.N. Basov \textit{et al}., Phys. Rev. Lett.
{\bf 74}, 598 (1995).
\bibitem{turneaure01} S.J. Turneaure, E.R. Ulm, and T.R.
Lemberger, J. Appl. Phys. {\bf 79}, 4221 (1996); S.J. Turneaure,
A.A. Pesetski, and T.R. Lemberger, \textit{ibid}. {\bf 83}, 4334
(1998).
\bibitem{note2} Reproducibility is much improved when all samples are made
with the same deposition procedure \cite{skinta}. For comparison,
there are similarly large variations in $\lambda ^{-2}(0)$ for
bulk LSCO, with single crystals having a substantially smaller
value than for powders \cite{paget}.
\bibitem{skinta} J.A. Skinta, M.-S. Kim, T.R. Lemberger, T.
Greibe, and M. Naito, to be published.
\bibitem{paget} K.M. Paget \textit{et al}., Phys. Rev. B {\bf 59}, 641
(1999), and references therein.

\end{references}
\end{document}